\documentclass[12pt]{article}
\usepackage{a4wide}
\usepackage{amsmath,amssymb,amsbsy}
\usepackage{latexsym,layout,amsfonts,amssymb,fontenc,xcolor,amsthm,multirow,amsfonts,bm,textcomp,epstopdf}
\usepackage{multirow}
\usepackage{slashbox}
\newsavebox{\uuunit}
\sbox{\uuunit}
    {\setlength{\unitlength}{0.825em}
     \begin{picture}(0.6,0.7)
        \thinlines
        \put(0,0){\line(1,0){0.5}}
        \put(0.15,0){\line(0,1){0.7}}
        \put(0.35,0){\line(0,1){0.8}}
       \multiput(0.3,0.8)(-0.04,-0.02){12}{\rule{0.5pt}{0.5pt}}
     \end {picture}}

\def\2{\frac12}
\def\4{\frac14}

\newcommand{\be}{\begin{equation}}
\newcommand{\ee}{\end{equation}}
\newcommand{\bea}{\begin{eqnarray}}
\newcommand{\eea}{\end{eqnarray}}

\begin{document}

\begin{titlepage}
\begin{center}

\hfill ROM2F/2013/15 \\
\hfill UG-2013-56

\vskip 1.5cm

{\Large \bf Heterotic-Type II duality and wrapping rules}

\vskip 1cm

{\bf E.A.~Bergshoeff\,$^1$, C. Condeescu\,$^{2,3}$, G. Pradisi\,$^{4,2}$ and F. Riccioni\,$^5$}

\vskip 25pt

{\em $^1$ \hskip -.1truecm Centre for Theoretical Physics,
University of Groningen, \\ Nijenborgh 4, 9747 AG Groningen, The
Netherlands \vskip 15pt }

{\em $^2$ \hskip -.1truecm INFN, Sezione di Roma ``Tor Vergata''\\ Via della Ricerca Scientifica 1, 00133 Roma, Italy
 \vskip 15pt }

{\em $^3$ \hskip -.1truecm
Department of Theoretical Physics, IFIN-HH\\
Str. Reactorului 30, 077125, Magurele-Bucharest, Romania
\vskip 15pt }

{\em $^4$ \hskip -.1truecm Dipartimento di Fisica, Universit\`a di Roma ``Tor Vergata''\\ Via della Ricerca Scientifica 1, 00133 Roma, Italy
 \vskip 15pt }

{\em $^5$ \hskip -.1truecm
 INFN Sezione di Roma,  Dipartimento di Fisica, Universit\`a di Roma ``La Sapienza'',\\ Piazzale Aldo Moro 2, 00185 Roma, Italy
 \vskip 15pt }

{email addresses: {\tt E.A.Bergshoeff@rug.nl}, {\tt Cezar.Condeescu@roma2.infn.it}, {\tt  Gianfranco.Pradisi@roma2.infn.it},  {\tt Fabio.Riccioni@roma1.infn.it}} \\

\end{center}

\vskip 0.5cm

\begin{center} {\bf ABSTRACT}\\[3ex]
\end{center}
We show how the brane wrapping rules, recently discovered in closed oriented string theories compactified on tori, are extended to the case of the Type IIA string compactified on K3. To this aim, a crucial role is played by the duality between this theory and the Heterotic string compactified on a four-dimensional torus $T^4$. We first show how the wrapping rules are applied to the $T^4/\mathbb{Z}_N$ orbifold limits of K3 by relating the D0 branes, obtained as D2 branes wrapping two-cycles, to the perturbative BPS states of the Heterotic theory on $T^4$.  The wrapping rules are then extended to the solitonic branes of the Type IIA string, finding agreement with the analogous Heterotic states. Finally, the geometric Type IIA orbifolds are mapped, via T-duality, to non-geometric Type IIB orbifolds, where the wrapping rules are also at work and consistent with string dualities.

\end{titlepage}

\newpage
\setcounter{page}{1}

\tableofcontents

\newpage

\setcounter{page}{1} \numberwithin{equation}{section}

\section{Introduction}

Most of the progress made in our understanding of non-perturbative string dualities is based on the study of BPS states. Indeed, for such  states the mass (or tension) can be determined exactly and does not receive quantum corrections \cite{Witten:1978mh}.
Elementary BPS  string states, oscillation modes of the fundamental strings, are explicitly present in the string spectra, while the existence of BPS Dirichlet branes can be inferred from the Ramond-Ramond sector of the spectra of Type II strings or, directly, from the spectra of open strings \cite{Polchinski:1998rq,Angelantonj:2002ct}. On the other hand, the existence of
additional non-perturbative BPS branes can only be predicted analyzing the low-energy supergravity theory, where they
manifest themselves through the corresponding superalgebras \cite{Townsend:1996xj}.  Moreover, their semiclassical approximations
arise as classical solitonic solutions preserving a corresponding portion of supersymmetry. These solutions transform in given representations of the global symmetry groups of the supergravity theories and, in general, the fact that the discretization of these groups correspond to string theory dualities dictates how a brane of a given string theory is mapped to a dual brane of the dual theory \cite{Hull:1994ys}.
In this paper we shall only deal with 1/2-BPS branes, preserving the largest possible amount of supersymmetry.

Recently, a full classification  of all 1/2-BPS branes of Type
IIA/IIB string theory compactified on a torus has been obtained \cite{Bergshoeff:2011qk,Kleinschmidt:2011vu,Bergshoeff:2012ex}. A crucial ingredient of this
classification are the properties of the various representations of the brane charges under the global U-duality symmetry group of the low energy supergravity. Denoting with
$d$ the number of internal directions, this symmetry group contains $\text{SO}(d,d) \times \mathbb{R}^+$, where  $\mathbb{R}^+$ is the string dilaton scaling and $\text{SO}(d,d)$ is the part of the symmetry whose discrete counterpart is T-duality.  Since $\text{SO}(d,d)$ does not affect the dilaton, one then obtains a classification of 1/2-BPS branes in terms of the dilaton weight, that is for a given scaling $\alpha$ of the brane tension $T$ with respect to the string coupling $g_S$ in the string frame:
\begin{equation}
\text{T}\ \sim \ g_S^\alpha \ . \label{alphascaling}
\end{equation}
While the fundamental branes have $\alpha=0$ and the D-branes have $\alpha=-1$, from this analysis one also obtains all the non-perturbative branes corresponding to larger negative integer values of $\alpha$.

A remarkable feature of the resulting classification is that one finds
constraints on the brane charges transforming in $\text{SO}(d,d)$
representations that are dual to the representations of the supergravity potentials. This result is well known for the branes that are electrically or magnetically charged with respect to the ``standard'' gauge potentials, {\it i.e.} those carrying on-shell degrees of freedom in the supergravity multiplet. Expressed in a group-theoretical language, the constraint is the highest-weight constraint of the corresponding representation.
The results of \cite{Bergshoeff:2011qk,Kleinschmidt:2011vu,Bergshoeff:2012ex} show that the same constraint occurs for the branes of the theory that have 2, 1 or 0 transverse directions and thus do not correspond to the standard gauge potentials of supergravity.
Such branes, called\footnote{As opposed to the standard branes, that have codimension greater than or equal to 3.} ``non-standard'' in \cite{Bergshoeff:2011mh}, are electrically charged with respect to $D-2$, $D-1$ and $D$-form potentials. The $D-2$-form potentials are magnetically dual to the scalars, which means that they also describe physical degrees of freedom although, differently from the lower-rank potentials, the duality relation with the scalars implies that some combinations of their field strengths vanish identically. The $D-1$-form potentials do not carry any degree of freedom, and their field strengths are dual to gauge coupling constants or mass parameters that occur in gauged supergravities. Finally, the field strengths of the $D$-form potentials vanish identically. The symmetry-group representations of the $D-1$ and $D$-form potentials
can be determined from the  requirement that the supersymmetry algebra and the gauge algebra close on them \cite{Bergshoeff:2005ac,Bergshoeff:2006qw,Riccioni:2007au,Bergshoeff:2007qi}.

For representations of $\text{SO}(d,d)$ the highest-weight constraint is solved by all the components satisfying the so-called ``light-cone rules'' \cite{Bergshoeff:2011zk,Bergshoeff:2011ee}. As we shall review in Section 2, they give a natural way of counting the number of independent 1/2-BPS branes.
Remarkably, for the branes with $\alpha=0,-1,-2$ and $-3$ this number can also be derived by applying the ``wrapping rules'' \cite{Bergshoeff:2011mh,Bergshoeff:2011ee}
\begin{align}
\alpha=0 \ : \quad  & \begin{cases} {\rm wrapped}   \ \rightarrow\   \ {\rm doubled}\\ {\rm unwrapped} \  \rightarrow \  {\rm undoubled} \ ,\end{cases} \nonumber\\~\nonumber\\
\alpha=-1 \ :\quad & \begin{cases} {\rm wrapped}   \ \rightarrow\   \ {\rm undoubled}\\ {\rm unwrapped} \  \rightarrow \  {\rm undoubled} \ ,\end{cases} \nonumber\\~\label{allwrappingrulesinonego}\\
\alpha=-2 \ :\quad  & \begin{cases} {\rm wrapped}   \ \rightarrow\   \ {\rm undoubled}\\ {\rm unwrapped} \  \rightarrow \  {\rm doubled} \ ,\end{cases} \nonumber\\~\nonumber\\
\alpha=-3 \ :\quad & \begin{cases} {\rm wrapped}   \ \rightarrow\   \ {\rm doubled}\\ {\rm unwrapped} \  \rightarrow \  {\rm doubled} \ ,\end{cases}\nonumber
\end{align}
\noindent that allow to find the number of branes in a given dimension once the branes in one higher dimension are known. This implies that the number of branes in any dimension can be obtained starting from the branes in 10 dimensions\footnote{The fact that there are two theories in ten dimensions implies that the wrapping rules when reducing from 10 to 9 dimensions are more subtle. Indeed, in this case ``doubled'' means that one has to consider both the IIA branes and the IIB branes, while ``undoubled'' means that one has to consider the 9-dimensional branes as coming either from IIA or from IIB, and by consistency both choices give the same result.}. As an example, let us consider the $\alpha=0$ rule.
Reducing the 10-dimensional theory to 9 dimensions in a direction longitudinal to the fundamental string F1, {\it i.e.}  when the fundamental string is wrapped, one gets a pair of 0-branes given by the F1 itself and the dimensionally reduced pp-wave. In this sense, the fundamental string doubles when it wraps. On the other hand, in 9 dimensions there is still a single fundamental string, because when the 10-dimensional theory is reduced in a direction transverse to the F1, {\it i.e.} when the fundamental string is unwrapped, the fundamental string itself is not doubled. This procedure can then be iterated and in $D=10-d$ dimensions gives $2d$ F0-branes and one F1-brane, which are exactly the numbers given by the light-cone rules.
Similarly, the $\alpha=-2$ NS5-brane in 10-dimensions gives a single NS4-brane in 9-dimensions (the dual of the fundamental string), while two 5-branes are present in 9 dimensions, one being the unwrapped NS5-brane, the other coming from the 10-dimensional KK6-monopole. That is, the NS5-brane doubles when it does not wrap. Again, this can be iterated to any dimension.

Besides the fundamental string (with $\alpha=0$) and its magnetic dual, the ($\alpha=-2$) NS5-brane, Type IIA and Type IIB strings have Dirichlet branes with $\alpha=-1$ and $p$ even or odd, respectively. The number of such branes in lower dimensions is then obtained by applying the $\alpha=-1$ wrapping rule, which implies that the D-branes are always undoubled, and thus their number is never increased by the contributions of generalized KK monopoles.
Finally, the Type IIB theory also contains a 7-brane with $\alpha=-3$, required to be the S-dual of the D7-brane, and a 9-brane with $\alpha=-4$, S-dual to the D9-brane. The $\alpha =-3$ branes in any dimensions are obtained by the related wrapping rule in \eqref{allwrappingrulesinonego}.  The situation is different for the $\alpha=-4$ branes.  In 7 dimensions and below, $\alpha=-4$ branes appear that cannot be obtained by wrapping rules.  However, there exists a specific irreducible $\text{SO}(d,d)$ representation of $\alpha=-4$ 1/2-supersymmetric space-filling branes whose number, in any dimensions, is determined by the Type IIB ($\alpha=-4$) 9-brane by means of the additional wrapping rule\footnote{This brane is space-filling and thus can only wrap.}
\begin{equation}
\alpha=-4 \ : \quad  {\rm wrapped}   \ \rightarrow\   \ {\rm doubled}  	\quad .\label{alpha=-4wrappingrules}
\end{equation}
The outcome of this whole analysis is that {\it all} the branes of Type IIA and Type IIB strings, when dimensionally reduced, satisfy specific wrapping rules.

In \cite{Bergshoeff:2012jb} the analogue of this  analysis  was performed for  the 1/2-BPS branes of the Heterotic string compactified on a torus $T^d$. The  T-duality symmetry in this case is $\text{SO}(d,d+16; \mathbb{Z})$ and the low energy theory, at least at generic points in the moduli space where Wilson lines break the gauge group to $\text{U}(1)^{2d+16}$, has a global symmetry  $ \text{SO}(d,d+16)\times\mathbb{R}^+ $, where $\mathbb{R}^+$ is again the string dilaton scaling. Thus, one can classify the brane charges as representations of $\text{SO}(d,d+16)$ for each value of $\alpha$, that in the Heterotic case is only even.  In \cite{Bergshoeff:2012jb} the single-brane states that preserve 1/2-supersymmetry were classified as in the maximal theory. Again, one finds that such branes satisfy the highest-weight constraint, and their number is determined by applying the light-cone rules to $\text{SO}(d,d+16)$.
The 1/2-BPS branes of the Heterotic string in 10 dimensions are the fundamental string F1 (with $\alpha=0$) and its dual NS5-brane (with $\alpha=-2$). The number of branes that one gets in $10-d$ dimensions by applying the light-cone rules again coincides with that of the single 1/2-BPS branes with $\alpha=0$ and $\alpha=-2$ in $10-d$, obtained by applying to the ten-dimensional theory the first and the third wrapping rules in \eqref{allwrappingrulesinonego}.
To summarize, {\it all} the branes of the ten-dimensional closed oriented string theories, upon toroidal compactifications, must satisfy specific wrapping rules
in order to give rise to the correct number of lower-dimensional supersymmetric branes.

The Heterotic string compactified on $T^4$ is conjectured to be dual to the Type IIA theory compactified on a K3 manifold \cite{Hull:1994ys,Witten:1995ex}. One can thus use the duality and the classification of the branes of the heterotic theory on the torus to gain some understanding of how the mentioned wrapping rules generalize to Type IIA string theory compactified on K3. This is the aim of this paper. In particular, we shall consider points in the moduli space of K3 corresponding to $T^4/\mathbb{Z}_N$ orbifolds, where the perturbative Type IIA string theory admits a CFT description. In general, if the K3 surface is non-singular, the Type IIA low-energy theory corresponds to a six-dimensional ${\cal N}=(1,1)$ supergravity coupled to 20 abelian vector multiplets. As we saw above, the dual Heterotic theory must be at a generic point in moduli space where all the vectors are abelian.  Non-abelian symmetries on the Heterotic side correspond to orbifold singularities of K3, with smooth cycles collapsing to zero size. The orbifold $T^4/\mathbb{Z}_N$ CFT descriptions of Type IIA on K3, however, do not correspond to these symmetry enhancement points, because there is a hidden non-vanishing $B$ field at every collapsed 2-cycle that renders the corresponding compactification not equivalent to a truly singular configuration \cite{Aspinwall:1995zi}, even though the corresponding K3 exhibits conical singularities.

A preliminary analysis of the wrapping rules of the Type IIA theory on the orbifold $T^4/\mathbb{Z}_2$ was performed in \cite{Bergshoeff:2012jb}. The idea of that paper was to extend the wrapping rules on the torus to the wrapping rules on the so called ``bulk cycles'' of the orbifold, {\it i.e.} those cycles of the torus that survive the orbifold projection. In the particular case of the $T^4/\mathbb{Z}_2$ orbifold, all the six 2-cycles of the torus are bulk 2-cycles, and applying the wrapping rules on them gives rise to the correct answer.  In this paper we clarify the analysis of \cite{Bergshoeff:2012jb} and generalize it to the other orbifold limits of K3. This is done by considering in particular the D0-branes of the Type IIA theory on the orbifold, that are dual to the perturbative 0-branes of the Heterotic theory on $T^4$. We find that to correctly account for the single brane states on the Type IIA side one has to identify also the ``fractional branes'', {\it i.e.} branes that wrap the so-called ``exceptional cycles'', zero-size (singular) collapsed cycles of the orbifold. The branes that only wrap the bulk cycles are those particular combinations of fractional branes with vanishing twisted R-R charge. Once a basis of fundamental homology cycles has been identified, one can check that the wrapping rules are satisfied due to the structure of the corresponding lattice. This can be done for any orbifold. We then apply the wrapping rules on these cycles also to the other branes of the IIA theory.

Only branes with $\alpha=0,-1$ and $-2$ are contained in the Type IIA theory in 10 dimensions.  As a consequence, only branes with the same values of $\alpha$ can be obtained in lower dimensions. On the other hand, the Type IIA theory is T-dual to the Type IIB theory. Implementing T-duality within the orbifold construction, one can think of the Type IIA orbifolds as equivalent to non-geometric orbifolds of the Type IIB theory.  As far as D-branes are concerned, T-duality maps a wrapped direction of a D-brane in one theory to a transverse direction of a D-brane in the dual theory. Therefeore, applying T-duality to K3 orbifold limits probes the non-geometric nature of the Type IIB orbifold, because it formally maps D-branes wrapped on even cycles on the IIA side to D-branes wrapped on odd cycles on the IIB side.  We show that one can obtain the D-branes of the orbifold theory by applying the $\alpha=-1$ wrapping rules of equation  \eqref{allwrappingrulesinonego} to the D-branes of Type IIB in 10 dimensions wrapping odd (non-geometric) cycles.  On the other hand, for $\alpha=0$ branes T-duality simply exchanges the two 0-branes in lower dimensions, namely it maps the fundamental string of one theory to the pp-wave of the other.  In this sense, they are ``blind'' to T-duality and only see (geometric) even cycles.  The rules persist for the other values of $\alpha$, including the $\alpha=-3$ and $\alpha=-4$ cases, not existing in the Type IIA theory.  Lower dimensional branes of Type IIB compactified on the T-dual (non-geometric) K3 are obtained applying the wrapping rules \eqref{allwrappingrulesinonego} to even cycles for even $\alpha$'s and to odd cycles for odd $\alpha$'s.  All these branes have corresponding dual branes in the Heterotic string compactified on $T^4$.

The plan of the paper is as follows. In section 2 we discuss the 1/2-BPS branes of the Heterotic string on $T^4$ and the resulting wrapping rules.
We first consider  perturbative states (subsection 2.1), and next the non-perturbative ones (subsection 2.2).  In section 3 we analyze the branes of the Type IIA string theory compactified on K3, showing how the duality between this theory and the Heterotic string on $T^4$ leads to a characterization of the K3 wrapping rules. In particular, in subsection 3.1 we review the Heterotic-Type IIA duality.  In subsection 3.2 we discuss the
D0-branes of the Type IIA string on K3 in the orbifold limits mapped, by string duality, to the perturbative 0-branes of the Heterotic theory.
In subsection 3.3 we review how T-duality maps the Type IIA string on K3 to a non-geometric compactification of the Type IIB string.
In subsection 3.4 we extend our analysis to Type II branes with $\alpha < -1$, showing how string-string duality with the Heterotic
theory is consistent with the wrapping rules both for the geometric Type IIA and the non-geometric Type IIB strings.  In the matching of branes,
an important role is played by ${\rm SO}(4,4)$ triality. Finally, in section 4 we present our conclusions and discuss some future directions.

\section{Heterotic string on $T^4$ and wrapping rules}

In this section we review the classification of 1/2-supersymmetric BPS branes in the Heterotic string compactified on a torus.   As a first step, it is
useful to repeat how the same classification was performed for the 1/2-BPS branes of the maximal theories in terms of
representations of $\text{SO}(d,d)$.

One way to classify the branes of the maximal theories  is to require that the coupling of the supergravity potentials to the
candidate 1/2-BPS branes occurs via a ``supersymmetric'' Wess-Zumino
(WZ) term\footnote{By ``supersymmetric'' we mean, in this context, that a
gauge-invariant Wess-Zumino term can be written down using only
world-volume fields that fit into a supermultiplet with 16
supercharges.}. This requirement is sufficient to find a full classification of the 1/2-BPS branes. The outcome of this analysis implies that all these branes carry charges satisfying the highest-weight constraint of the representation of the global symmetry group. It means that the number of components of the representation that satisfy the constraint coincides with the number of weights that can be chosen to be highest weights. In a weight diagram, they are always the longest weights of the representation \cite{Bergshoeff:2013sxa}.

As an example, we can consider the fundamental 0-branes associated to the supergravity 1-form potentials $A_{1,A}$, with charge $Q_A$ in the vector representation of $\text{SO}(d,d)$.  The supersymmetry requirement implies the constraint
\begin{equation}\label{condition}
Q^A Q_A=0\,.
\end{equation}
Therefore, the 1/2-BPS fundamental 0-branes correspond to the $2d$ light-like directions of $\text{SO}(d,d)$.
This phenomenon comes out to be completely general.  Indeed, by choosing a basis of light-like coordinates for $\text{SO}(d,d)$, that we may denote
\begin{equation}
i \pm  = x_i \pm t_i \qquad i =1,...,d \ ,
\end{equation}
with $x_i$ space-like and $t_i$ time-like, the number of longest weights of a $\text{SO}(d,d)$ representation is given by the number of components satisfying
 the following ``light-cone rules'' \cite{Bergshoeff:2011zk,Bergshoeff:2011ee}:

\begin{itemize}

\item For a representation with $n$ antisymmetric indices, the light-cone rule selects the components $i_1\pm \ i_2\pm \ \ ... \ \ i_n\pm$ antisymmetric in the $i$'s (regardless of the $+$ or  $-$ choice), which selects ${d \choose n} \times 2^n$ components inside a representation of dimension ${2d \choose n}$;

\item for representations with two sets of antisymmetric indices, {\it i.e.} representations whose Young tableaux are given by two columns, one with $m$ boxes and the other with $n$ boxes and $m \geq n$, on top of the rule above one also has to impose that the $n$ indices have to be the same as $n$ of the $m$ indices;

\item for tensor-spinor representations, the number of longest weights is given by ${d \choose n} \times 2^{d-1}$, where $n$ is the number of antisymmetric vector indices and $2^{d-1}$ is the dimension of the chiral spinor representation of  $\text{SO}(d,d)$.

\end{itemize}
The charges of the branes of the Type-II theory compactified on a torus all belong to representations that fit in the three classes above\footnote{In particular they are ``bosonic'' for  $\alpha$ even and ``fermionic'' for $\alpha$ odd, which means that the number of 1/2-BPS branes are obtained using the first two rules for $\alpha$ even and the third rule for $\alpha$ odd.}. These representations can for instance be found in \cite{Bergshoeff:2012ex} for all the different values of $\alpha$, and from them one obtains the number of all 1/2-BPS branes in any dimension using the rules above.

We now move to consider the branes of the Heterotic theory in $D=10-d$ dimensions. As we shall review in more detail in Section $3$, at a generic point in the moduli space this theory has a  gauge symmetry $\text{U}(1)^{2d+16}$, and the low-energy theory possesses a global symmetry $\text{SO}(d,d+16)$ which is broken to the T-duality symmetry $\text{SO}(d,d+16; \mathbb{Z})$ in the quantum theory.
The scalars parameterize the manifold $\mathbb{R}^+ \times \text{SO}(d,d+16)/ [\text{SO}(d) \times \text{SO}(d+16)]$, where $\mathbb{R}^+$ is the scaling corresponding to the $D$-dimensional dilaton.
In \cite{Bergshoeff:2012jb} the 1/2-BPS branes of this theory were classified according to their dilaton scaling, as indicated in \eqref{alphascaling}, with
$\alpha$ a non-positive even integer, as opposed to the maximal case where $\alpha$ can be even or odd. The single-brane states
correspond again to the independent components of the charges that satisfy the highest-weight constraint.
Exactly as in the maximal case, the number of independent components of the representation that satisfy this constraint precisely coincides with the number of components selected by the light-cone rules\footnote{Only the bosonic light-cone rules are needed because $\alpha$ is even.}.  In particular, the branes with $\alpha=0$ and $\alpha=-2$ can be derived from the branes of the ten-dimensional theory using  exactly the first and the third wrapping rules in eq. \eqref{allwrappingrulesinonego}.

The single-brane states classified in \cite{Bergshoeff:2012jb} are such that  bound states of them can  still preserve 1/2 of the  supersymmetry.  For instance, for the $\alpha=0$ 0-branes,  with charge $Q_A$ in the vector representation of $ \text{SO}(d,d+16)$, the rule gives $2d$ light-like directions, corresponding to $Q^2=0$. In the Heterotic string, they are the purely momentum or purely winding states.  On the other hand, there exists a tower of 1/2-BPS perturbative states with $Q^2\neq 0$. They are interpreted as bound states of the single brane states.  It should be observed that a bound state can preserve the same amount of supersymmetry as a single-brane state due to the fact that the BPS condition is {\it degenerate}. This is precisely what happens also in the maximal theory as far as the non-standard branes are concerned: for instance, the two 7-branes of the Type IIB theory mentioned earlier have a degenerate BPS condition, and one can indeed find 7-brane solutions preserving 1/2 of the supersymmetry and corresponding to bound states of the two single 7-branes \cite{Greene:1989ya,Bergshoeff:2006jj}. In the Heterotic theory this feature is completely general and applies to all the branes, not only to the non-standard ones.  In order to show this phenomenon in detail, it is useful to analyze the central charges of the supersymmetry algebra. In view of the duality with the Type IIA theory compactified on K3, we focus in particular on the six-dimensional case, corresponding to the Heterotic theory on $T^4$. The R-symmetry of the six-dimensional theory is USp(2)$\times$USp(2), and the central charges (including the momentum operator) are
   \begin{equation}
   Z_{a \dot{a}} \ ({\bf 2},{\bf 2}) \quad P_{\mu} \ ({\bf 1},{\bf 1}) \quad Z_\mu \ ({\bf 1},{\bf 1})\quad Z_{\mu \nu, a \dot{a}} \
  ({\bf 2},{\bf 2}) \quad Z_{\mu\nu\rho, ab}^+ \ ({\bf 3},{\bf 1}) \quad Z_{\mu\nu\rho, \dot{a}\dot{b}}^- \ ({\bf 1},{\bf 3}) \ ,\label{centralchargesD=6}
  \end{equation}
where $\mu,\nu,...$ are space-time indices, the $+$ and $-$ denote space-time (anti)self-duality and $a,b$ ($\dot{a},\dot{b}$) are indices of the doublet of the first (the second) USp(2).  We have explicitly indicated the USp(2)$\times$USp(2) representations for each charge.
As it will emerge in the rest of the section, all these central charges are degenerate, {\it i.e.} in all cases there is more than one brane associated to a given central charge component.  In the next subsections we shall review  the results of \cite{Bergshoeff:2012jb}, giving the single-brane states and showing how the number of $\alpha=0$ and $\alpha=-2$ branes is obtained using the wrapping rules. We shall also determine the degeneracy of the central charges, focusing in particular on the six-dimensional case. We shall start first by considering perturbative string states and then analysing the cases with $\alpha  < 0$, which correspond to non-perturbative states in string theory.

\subsection{Perturbative BPS states}

Before discussing the perturbative BPS states in the Heterotic theory on the torus, it is instructive to consider BPS states in the Type-II theory compactified on the same torus \cite{Dabholkar:1995nc}. A set of states with a given momentum $M$  and winding $W$ group together to form a short multiplet if there are no oscillators in either the left or the right sector. Such states are 1/4-BPS, and in the particular case of a 1-dimensional torus, {\it i.e.} a circle, they form an $\text{SO}(1,1;\mathbb{Z})$ lattice with
invariant $Q^2 = MW$. The degenerate case in which there are no oscillators in both the left and the right sectors leads to 1/2-BPS states. Such states have $Q^2 =0$ and therefore they are either purely momentum or purely winding states. One can then interpret the 1/4-BPS states, with both $M$ and $W$ non-zero, as bound states of the 1/2-BPS states.
This naturally generalizes to a compactification on $T^d$, where now the momentum and winding states form an $\text{SO}(d,d;\mathbb{Z})$ lattice. The 1/2-BPS states  have $Q^2=0$, and one can always choose a
basis in which such states are either purely momentum or purely winding states.  Again, the 1/4-BPS states with non-zero $Q^2$ are bound states of them.

In the Heterotic theory the picture is different, because only the left sector is supersymmetric and any state with no left-moving oscillators is a 1/2-BPS state, regardless of the value of $Q^2$ \cite{Schwarz:1993mg, Dabholkar:1990yf}.  Thus, a bound state of a purely momentum and purely winding state keeps preserving the same amount of supersymmetry. The charge in the Heterotic case belongs to the $\text{SO}(d,d+16;\mathbb{Z})$ lattice.
In the low-energy theory, these states correspond to BPS black-hole solutions of the effective action  \cite{Callan:1995hn,Dabholkar:1995nc}.

Let us show how these results are embedded in the general classification of 1/2-BPS branes in maximal and half-maximal supergravity. In the maximal case, the degeneracy of the central charges is always 1 for standard branes, while it is higher than 1 for non-standard branes \cite{Bergshoeff:2013sxa}. Thus, in any dimension higher than 3 there is a one-to-one correspondence between 1/2-supersymmetric 0-branes and components of the corresponding central charges. The $\alpha=0$ 0-branes of maximal supergravity theories are characterized by a charge $Q_A$ in the vector representation of  $\text{SO}(d,d)$, and the condition for the brane to be 1/2-supersymmetric is the highest-weight constraint $Q^2=0$. The fact that the central charges are not degenerate implies that different light-like directions never give 1/2-supersymmetric branes preserving all the same
supersymmetries. A supersymmetric bound state of these branes, with $Q^2 \neq 0$, is thus a 1/4-BPS state, exactly as mentioned above.

The picture changes for all the 1/2-supersymmetric branes of the  half-maximal theories (as well as for non-standard branes of the maximal theories). In this case the degeneracy of the central charge implies that two different single 1/2-BPS branes can preserve the same supercharge components giving rise to a bound state that is still 1/2-BPS. As an example, let us consider the 0-branes of the Heterotic theory in six dimensions. They are $\alpha=0$ branes whose charge $Q_A$ belongs to the vector representation of $\text{SO}(4,20)$. The single-brane constraint in this case is $Q^2=0$, and the independent branes correspond to the eight different light-like directions that one can pick out inside $\text{SO}(4,20)$. From eq. \eqref{centralchargesD=6}, one can see that there are 4 central charges, corresponding to the ${\bf (2,2)}$ of the R-symmetry group, with a corresponding degeneracy equal to 2. Denoting the light-like directions of $\text{SO}(4,20)$ as $i\pm$, with $i =1,...,4$, the charges $Q_{i+}$ and $Q_{i-}$ correspond to the same central charge. Consequently, the corresponding brane configurations preserve the same supercharges and their bound state is still 1/2-BPS. This reproduces the string theory result mentioned before in the context of the Heterotic theory.

Together with the $\alpha=0$ 0-branes, the Heterotic theory in any dimension clearly possesses a single $\alpha=0$ 1/2-BPS 1-brane, which is the string itself.
In \cite{Bergshoeff:2012jb} it was shown that all the 1/2-BPS $\alpha=0$ single-brane states of the Heterotic theory can be obtained from the ten-dimensional string by applying the $\alpha=0$ wrapping rule of eq. \eqref{allwrappingrulesinonego}
on each compactified direction, starting from a single 1-brane in ten dimensions. This indeed leads to $2d$ independent single 0-brane states, corresponding to the different light-like directions that one can choose in $\text{SO}(d,d+16)$.

\subsection{Non-perturbative BPS branes}

Besides the perturbative BPS states, the Heterotic theory in six dimensions also contains non-perturbative BPS states with $\alpha=-2$ and $\alpha=-4$.  One classifies these branes in terms of the $\text{SO}(4,20)$ representation of the corresponding charge. The $\alpha=-2$ $p$-branes have charges $Q_{A_1 ...A_{p-1}}$ in the antisymmetric representation with $p-1$ indices. Exactly as for the fundamental 0-branes, the charge of the 1/2-supersymmetric single brane states satisfies the highest-weight constraint of the representation, namely their charges have indices along the light-like directions, $Q_{i_1 \pm ... i_{p-1} \pm}$, with the $i$'s all different. Counting the independent components, this leads to 1 1-brane, 8 2-branes, 24 3-branes, 32 4-branes and 16 5-branes. The 1-brane and the 2-branes are the magnetic duals of the fundamental string and of the perturbative  0-branes, respectively.

The $\alpha=-4$ branes include a 4-brane with charge $Q_A$ and a 5-brane with charge $Q_{AB}$ symmetric in $AB$. The constraint on the charge for the 4-branes is $Q^2=0$, exactly  as for the perturbative 0-branes, while the constraint for the 5-branes imposes not only that the indices have to be along the light-like directions, but also parallel. In both cases, it leads to 8 independent branes. The complete result is summarized in Table \ref{branesD=6heterotic}.

\begin{table}[h]
\begin{center}
\begin{tabular}{|c||c|c|c|c|c|c|}
\hline \rule[-1mm]{0mm}{6mm} \backslashbox{$\alpha$}{$p$} & $p=0$ & $p=1$ & $p=2$ & $p=3$ & $p=4$ &$p=5$ \\
\hline \hline \rule[-1mm]{0mm}{6mm} $\alpha=0$ & 8 & 1& & & & \\
\hline \rule[-1mm]{0mm}{6mm} $\alpha=-2$ & & 1&8&24&32&16\\
\hline \rule[-1mm]{0mm}{6mm} $\alpha=-4$ & & &&&8 &8\\
\hline
\end{tabular}
\caption{\sl \footnotesize The number of independent single-$p$-brane 1/2-BPS states in the Heterotic theory in six dimensions for the different values of $\alpha$.}
\label{branesD=6heterotic}
\end{center}
\end{table}

The analysis of the degeneracy of the central charges was performed in \cite{Bergshoeff:2012jb}. As already mentioned, the four $n=0$ BPS conditions correspond to the 8
$\alpha=0$ 0-branes and have degeneracy 2. The translation generator corresponds to the pp-wave and to the
fundamental string. The  $n=1$ central charge $Z_\mu$ in eq. \eqref{centralchargesD=6} corresponds to
the solitonic 1-brane and to the KK-monopole.
The four $n=2$ $Z_{\mu\nu , a \dot{a}}$  are related to the 8 solitonic 2-branes and therefore have
degeneracy 2.  The six $n=3$ central charges correspond to defect branes with degeneracy 4. The four $n=4$ central charges are the dual of $Z_{\mu\nu , a \dot{a}}$ and correspond to the 32 $\alpha=-2$  domain walls (with degeneracy 8) and to the 8 $\alpha=-4$ domain walls (with degeneracy 2). Finally, the charge $Z_\mu$ can be dualized to give an $n=5$ central charge which is a singlet of the R-symmetry. Thus, the degeneracy of the space-filling branes is 16 for $\alpha=-2$ and $8$ for $\alpha=-4$. As already discussed, bound states of branes with degenerate central charge are still 1/2-supersymmetric.

To conclude this section, we would like to stress that the number of independent $\alpha=-2$ single 1/2-supersymmetric branes can be obtained starting from the 10-dimensional NS5-brane and applying the solitonic ($\alpha=-2$) wrapping rule in eq. \eqref{allwrappingrulesinonego}.

\section{Heterotic-Type II duality and K3 wrapping rules}

The aim of this section is to make use of the conjectured duality between the Heterotic theory on $T^4$ and the Type IIA theory on K3 to derive wrapping rules for the branes reduced on K3. This analysis was initiated in \cite{Bergshoeff:2012jb} by considering the particular case of the orbifold $T^4/\mathbb{Z}_2$ and by showing that the duality is perfectly consistent with  extending the wrapping rules of the Type IIA branes on the torus to the wrapping rules on the inherited bulk cycles of the orbifold. Here, we generalize the extension to any orbifold limit of K3. Moreover, using T-duality between Type IIA  and Type IIB strings, we show that the branes of the Type IIA theory on K3 with $\alpha=0,-1,-2$ and $-3$ can all be obtained by applying the wrapping rules to the Type IIB branes reduced on non-geometric orbifolds, with the restriction that they wrap even cycles for even $\alpha$ and odd cycles for odd $\alpha$.

We start by first reviewing the Heterotic-Type IIA duality. Then we discuss the D0-branes of the Type IIA theory on a $T^4/ \mathbb{Z}_N$ orbifold, showing the natural choice of a basis of homology 2-cycles on which the wrapping rules can be applied. The map of D0-branes of Type IIA theory to the perturbative 0-branes of the Heterotic theory  is consistent with the string duality. Moreover, we review how the Type IIA orbifold can be viewed, using T-duality, as a non-geometric orbifold of the Type IIB theory. Finally, we extend the analysis to all the non-perturbative branes.

The mapping of the branes under Heterotic-Type II string duality
is encoded in the relation
\begin{equation}
  \alpha_{Het} =-\alpha_{II} - (p+1) \ ,\label{alphahetalphaIIA}
\end{equation}
where $\alpha_{Het}$ and $\alpha_{II}$ are the dilaton scalings of the tension of the corresponding $p$-branes in the two theories.  In particular,
the branes contained in Table \ref{branesD=6heterotic} are mapped to the branes of Type IIA on K3, listed in Table \ref{IIBnongeo}. It should be noticed that the rows of  Table \ref{branesD=6heterotic} precisely correspond to the diagonals of Table \ref{IIBnongeo}.

\begin{table}[h!]
\begin{center}
\begin{tabular}{|c||c|c|c|c|c|c|}
\hline \rule[-1mm]{0mm}{6mm} \backslashbox{$\alpha$}{$p$} & $p=0$ & $p=1$ & $p=2$ & $p=3$ & $p=4$ &$p=5$ \\
\hline \hline \rule[-1mm]{0mm}{6mm} $\alpha=0$ &  & 1& & & & \\
\hline \rule[-1mm]{0mm}{6mm} $\alpha=-1$ & 8 & & 8 && 8&  \\
\hline \rule[-1mm]{0mm}{6mm} $\alpha=-2$ & & 1&&24 & &8\\
\hline \rule[-1mm]{0mm}{6mm} $\alpha=-3$ & & && & 32&\\
\hline \rule[-1mm]{0mm}{6mm} $\alpha=-4$ & & && & &16\\
\hline
\end{tabular}
\caption{\sl \footnotesize  1/2-BPS branes of Type IIA theory compactified on K3.
\label{IIBnongeo}}
\end{center}
\end{table}

\subsection{Heterotic-Type IIA duality}
It has been conjectured that the Type IIA string compactified on K3 is S-dual to the Heterotic String on a four-torus \cite{Hull:1994ys, Witten:1995ex}.  There are several evidences of it, related mainly to the analysis of the low energy effective actions and to the comparison of moduli spaces and quantum corrections.  First of all, both theories have a low energy spectrum that exhibits an ${\cal N}=(1,1)$ (non-chiral) supergravity in six dimensions coupled to 20 abelian vector multiplets.  While this is unavoidable for the Type IIA, it requires to sit at a generic point in the moduli space of the Heterotic string.  Indeed, on the Heterotic side, the gauge group is always of the form $\text{U}(1)^4 \times G$, where $G$ contains a semisimple simply laced Lie group. The Wilson lines must be ``generic'' in order to sit at points where $G$ is reduced to the maximal abelian subgroup $\text{U}(1)^{16}$.

In order to discuss some aspects of the duality, it is convenient to briefly review the Narain \cite{Narain:1985jj,Narain:1986am,Giveon:1994fu} construction of the moduli space of the toroidal compactifications.  It can be described introducing a self-dual even Lorentzian lattice $\Gamma^{4,20}$ spanned by the vectors $(p_L, p_R)$, built with momenta and windings.  Each lattice can be generated by rotations in $\text{SO}(4,20)$ on a reference lattice or, equivalently, by suitably choosing the background parameters, the internal metric, the antisymmetric two-form, the dilaton and the Wilson lines, related to the vev's of the gauge fields in the Cartan subalgebra of the ten-dimensional group.  The spectra are invariant under independent rotations of the left and right movers.  As a result, the parameter  space is the coset
\begin{equation}
 \frac{\text{SO}(4,20)}{\text{SO}(4) \times \text{SO}(20)} \ , \label{backgspace}
 \end{equation}
of dimension 80. Clearly, not all the values of the parameters provide different models.  Indeed, it can be shown that a transformation in
$\text{SO}(4,20;\mathbb{Z})$ (the T-duality group) gives rise to the same lattice, resulting in a moduli space coincident with the Grassmannian
\begin{equation}
   {\cal M} \ = \ \text{SO}(4,20;\mathbb{Z}) \, \setminus \, \frac{\text{SO}(4,20)}{\text{SO}(4) \times \text{SO}(20)}   \ . \label{modspaone}
\end{equation}
Finally, we have to add the dilaton, whose vev's are parameterized by an arbitrary positive number.  This results into the following  moduli space:
\begin{equation}
   {\cal M}_{Het} \ = \ \text{SO}(4,20;\mathbb{Z}) \, \setminus \, \frac{\text{SO}(4,20)}{\text{SO}(4) \times \text{SO}(20)} \ \times {\mathbb{R}}^+ \ . \label{modspatwo}
\end{equation}

On the Type IIA side, on the other hand, the moduli space is directly linked to the (quantum) geometry of K3 \cite{Aspinwall:1994rg}.  The complex structure moduli space
is described by the possible periods and coincides\footnote{There are subtleties, irrelevant to us, with orientation, see \cite{Aspinwall:1994rg, Nahm:1999ps}.} with the space of the Einstein metrics of unit volume
\begin{equation}
   {\cal M}_{E} \ = \ \text{SO}(3,19;\mathbb{Z}) \, \setminus \, \frac{\text{SO}(3,19)}{\text{SO}(3) \times \text{SO}(19)}  \ . \label{modspathree}
\end{equation}
The (stringy) presence of the $B$-field in the non-linear sigma model of Type IIA on K3 has the effect of enlarging the moduli space
again to the factor of eq. (\ref{modspaone}).  Indeed, in the natural decomposition
\begin{equation}
 \frac{\text{SO}(4,20)}{\text{SO}(4) \times \text{SO}(20)} =  \frac{\text{SO}(3,19)}{\text{SO}(3) \times \text{SO}(19)} \times {\mathbb{R}}^+ \times {\mathbb{R}}^{3,19} \  \label{modspafour}
\end{equation}
the three factors correspond to the Einstein metrics of unit volume, the volume and the $B$-field, respectively.  It should be noticed that the $B$-field is an element of the $H^2(K3,\mathbb{R})$, and as such it is a $22$-component field, modulo integer translations that are elements of the (integer) cohomology $H^2(K3,\mathbb{Z})$ and reflect the symmetry of the $\sigma$-model path-integral. The modular group also factorizes, resulting into the $ SO(3,19;\mathbb{Z})$ related to the global diffeomorphisms of K3 that, together with the mentioned translations, reconstruct the $SO(4,20;\mathbb{Z})$.  Taking into account the dilaton, the moduli space of Type IIA on K3 is the same as the Heterotic string on $T^4$ of eq. (\ref{modspatwo}) \cite{Seiberg:1988pf,Aspinwall:1994rg}
\begin{equation}
   {\cal M}_{IIA} \ = \ \text{SO}(4,20;\mathbb{Z}) \, \setminus \, \frac{\text{SO}(4,20)}{\text{SO}(4) \times \text{SO}(20)} \ \times {\mathbb{R}}^+ \ . \label{modspafive}
\end{equation}

Of course, the fact that at a generic point in the moduli space of the Heterotic string the two theories exhibit the same low-energy spectrum as well as the
same moduli space is not sufficient to justify the conjectured S-duality.  Indeed, one should compare the BPS states of the two theories and also check the
string corrections on the two sides.  The duality does not mix the dilaton with the rest of the moduli space, so in view of the vastly different CFT's, the only
possibility is a strong-weak coupling duality, with $\phi_{Het}=-\phi_{IIA}$ and the $B$-field mapped to the dual $B$-field, as required by string-string duality.  Moreover, the perturbative states of one theory must correspond to non-perturbative states (solitons) in the other theory.  In \cite{Sen:1995cj} and \cite{Harvey:1995rn} it has been proved that a non singular soliton of the Type IIA string can be identified with the fundamental Heterotic string, admitting the correct 24-parameter moduli space.  Moreover, in the same reference \cite{Sen:1995cj}, a non-singular soliton solution of the Heterotic theory was constructed, that can be identified with the Type IIA fundamental string, since it does not allow for charged deformations.  These identifications can be traced back to a ten-dimensional rationale since, for instance, the fundamental Heterotic string in Type IIA can be interpreted as the IIA NS5 brane wrapping the whole K3.

Coming back to the enhancement points, the issue is that the perturbative Type IIA spectrum can only contain abelian gauge fields. Thus, the non abelian completions must come from non-perturbative (solitonic) sectors.  Witten \cite{Witten:1995ex} noticed that such kind of states should be
in general massive and can appear at the massless level if and only if the K3 exhibits quotient singularities, namely if the K3 is an orbifold.
On the other hand, orbifold singularities are known to be of a conical nature and to correspond to limit points in the moduli space where some smooth cycles collapse to zero-size.  The intersection matrices related to the collapsed cycles can be classified according to A-D-E type diagrams, identical to the Dynkin diagrams of the simply laced Lie algebras.  It is thus natural to conjecture that the enhancement configurations correspond to the orbifold limits in which the R-R vectors wrap collapsing cycles becoming massless in the zero-size limit \cite{Strominger:1995cz}.  It should be also observed, as stressed in \cite{Aspinwall:1995zi}, that the orbifold CFT is perfectly well defined and cannot describe non abelian vectors.  The reason is that, in the conventional (toroidal) orbifold CFT description of Type IIA on K3, there is a hidden non-vanishing $B$-field at every collapsed cycle that renders the corresponding K3 not properly equivalent to the singular orbifold configuration.  In other words, the points in the moduli space with enhancement of the gauge group are not accessible through an orbifold CFT description on the Type IIA side. They would give rise, genuinely, to a singular CFT and cannot be used to check the duality.  Other checks, however, are possible. For instance in \cite{Kiritsis:2000zi}, at least in the case in which the K3 is taken at the $T^4/\mathbb{Z}_2$-orbifold point, it has been verified that the BPS terms $Tr(F^4)$ in the twisted directions coincide.  The comparison is subtle, since on the Heterotic side the amplitude is a one-loop diagram, while on the Type IIA side it is
a tree level diagram of twist fields. A crucial ingredient in the identification is the $\text{SO}(4,4)$ triality relating vector to spinor and to conjugate-spinor representations.

\subsection{D-branes in Type IIA on K3}

As is well known \cite{Polchinski:1998rq}, D-branes in Type II theories are (BPS) solitons carrying Ramond-Ramond charge identical to a tension scaling like $g_S^{-1}$.  In the (static) weak coupling limit and in flat space, they can be described as topological defects (basically hyperplanes) on which the open strings can end.  More generically, considering curved internal target spaces, they correspond to the so called supersymmetric cycles they wrap, being pointlike in the remaining internal directions \cite{Ooguri:1996ck, Bershadsky:1995qy}. In this limit, the fluctuation of their shape are described as (tree level) exchanges of  closed strings that can be studied as dual channel amplitudes of (one loop) propagations of the corresponding open strings \cite{Angelantonj:2002ct}.  Microscopically, it is then natural to associate the presence of D-branes to boundary states \cite{Callan:1987px, Cardy:1989ir}. They can be characterized as combinations of the so-called Ishibashi states \cite{Ishibashi:1988kg}, a sort of generalized coherent states of closed string modes on the related Fock space.  They may also be linked to the characters of the irreducible representations of the superconformal algebra \cite{Cardy:1989ir, Cardy:1991tv, Pradisi:1996yd}.

The boundary states and the corresponding characters carry the geometrical informations related to the compactification manifold.  Unfortunately, in the general case a Conformal Field Theory description is not available.  So, for an explicit description, we have to resort to very simple regions in the moduli space of compactifications, like tori, orbifold limits of Calabi-Yau spaces or isolated special points, for instance those corresponding to Gepner models.  The latter are rational CFT, {\it i.e.} they exhibit a finite number of boundary states or, equivalently, a finite number of primary fields.  A given Calabi-Yau compactification is characterized by an ${\cal N}=(2,2)$ superconformal theory on the world-sheet in the case of CY threefolds, or an ${\cal N}=(4,4)$ in the K3 case.  Boundary states are defined giving ``gluing conditions'' that relate generators of the left and right superconformal algebras through the world-sheet parity operator action or its mirror relatives, obtained by combining it with an order two outer automorphism.  For a generic Calabi-Yau compactification, there are two possible choices preserving 1/2 of the superalgebras, called $A$ and $B$ boundary conditions and related precisely to the $A$- and $B$-type of D-branes.  For instance, in ten dimensions the D-branes of the T-dual Type IIA and Type IIB theories are the even (Type A) branes of the Type IIA and the odd (Type B) branes of the Type IIB.  In this case, of course, the automorphism can be seen as T-duality, after a compactification on a circle.

From the target space point of view, the BPS branes wrap supersymmetric cycles of the Calabi-Yau space \cite{Ooguri:1996ck, Bershadsky:1995qy}.  Mirror symmetry maps even cohomology of one space into odd cohomology of the mirror. This reflects itself in the fact that B boundary conditions are related to D-branes wrapping even cycles that are holomorphic submanifolds.  Indeed, locally, the D branes satisfy Dirichlet boundary conditions orthogonally to the cycle, allowing for a complex structure on the cycle itself.  On the other hand, A-type boundary conditions implies that the supersymmetric cycles, called SLag (Special Lagrangian cycles) must be middle-dimensional.  In the K3 case, of course, there are only even cycles, but the 2-cycles are also the middle-ones. Indeed, the $0$- and the $4$-cycles are genuinely holomorphic, while the boundary conditions that preserve ${\cal N}=1$ superconformal algebra are parametrized by an $\text{SO}(3)$ group that reflects the freedom in the choice of the complex structure.  This is consistent since, being K3 self-mirrored, the $2$ cycles can be interpreted as holomorphic submanifolds or SLag submanifolds, depending on the choice of the complex structure.

In the presence of enhancement points, a general phenomenon is the appearance of additional automorphisms.  For instance, when a CFT description is available with a symmetry ${\cal A} \times {\cal A}$, one may have boundary states that preserve the chiral algebra ${\cal A}$ \cite{Fuchs:1997kt, Fuchs:2000cm}. On tori or toroidal orbifolds, more options are possible, like
permutations or rotations.  The CFT, of course, needs also the specification of the modular invariant combination of the chiral and antichiral sectors that determines the boundary states  coupled to the bulk theory \cite{Pradisi:1996yd}.

In this paper, we are interested into understanding the wrapping rules related to D-branes.  In particular, we would like to support their conjectured validity for generic compactifications on K3, in the 1/2-maximal case.  As reviewed in the introduction (see the second wrapping rule in eq. \eqref{allwrappingrulesinonego}), the wrapping rules for D-branes are trivially one-to-one. We should argue that they derive solely from the analysis of the holomorphic cycles, namely we have to explicitly characterize the so-called Picard group and classify the resulting D-branes of Type IIA.  Of course, this is not doable. We can at most resort to simpler cases, like for instance the toroidal orbifolds, where the Picard group and more generally the integer (co)homology\,\footnote{Due to Poincar\'e duality, we often trade cycles with forms.} of K3 can be constructed explicitly.  D-branes on orbifolds are known, of course. We want to relate them to the wrapping rules. To this aim, it is useful to briefly review the $\mathbb{Z}_N$ orbifold compactification of Type IIA, with $N=2,3,4,6$ and the characterization of  supersymmetric branes wrapping the homology cycles of K3.  We want to stress that the $B$-field is counted as the two-form coupled to the fundamental strings.  For this reason only the classical and not the {\it quantum} geometry of K3 is relevant to us.  In particular, the important object is the integer (co)homology of K3 and its relation to supersymmetry: D-branes wrapping supersymmetric cycles are the objects we have to count on K3.

Let us start by reviewing the $\mathbb{Z}_N$-orbifold limits of K3 \cite{Walton:1987bu, Wendland:2000ry}.  The geometry is the one of $T^4/\mathbb{Z}_N$, where the four torus can be chosen to be $T^2\times T^2$.  Each complex two-torus has a fundamental cell with a $\mathbb{Z}_N$ symmetry and the quotient action on the complex coordinates $(z_1, z_2)$ is given by
\be
(z_1, z_2) \rightarrow (g z_1, g^{-1} z_2)\ , \quad g\in \mathbb{Z}_N \ .
\label{orbaction}
\ee
It reduces the volume by a factor of $1/N$ and gives rise to a set of fixed points corresponding to conical singularities.  This kind of singularities are described by an $A-D-E$ root lattice $\Gamma$.  In particular, for a $\mathbb{Z}_N$ singularity of order $k$, the lattice is the one of $A_{k-1}$.  The four cases are reported in Table \ref{lattices}.  The  $\mathbb{Z}_2$-orbifold has 16 points of order two, the $\mathbb{Z}_3$-orbifold 9 points of order three, the $\mathbb{Z}_4$-orbifold 4 points of order four and six (doublets) of order two and, finally, the $\mathbb{Z}_6$-orbifold exhibits 1 point of order six, 4 (doublets) of order three and 5 (triplets) of order two.
\begin{table}[h]
 \begin{center}
  \begin{tabular}{|c||c|c|c|c|}
\hline
~&~&~&~&~\\
 $G$ & $\mathbb{Z}_2$ & $\mathbb{Z}_3$ & $\mathbb{Z}_4$ & $\mathbb{Z}_6$\\
 ~&~&~&~&~\\
\hline
~&~&~&~&~\\
  $\Gamma$ & $\mathbb{A}_1^{16}$ & $\mathbb{A}_2^{9}$ & $\mathbb{A}_3^{4}\oplus \mathbb{A}_1^{6}$ & $\mathbb{A}_5\oplus \mathbb{A}_2^{4}\oplus \mathbb{A}_1^{5}$\\
  ~&~&~&~&~\\
\hline
  \end{tabular}
  \caption{\sl \footnotesize Fixed point lattices.}
\label{lattices}
 \end{center}
\end{table}
The orbifolds can be blown-up to K3 with the minimal resolution procedure.  A fixed point corresponds to a singularity locally isomorphic to
$\mathbb{C}^2/\mathbb{Z}_k$.  Each fixed point is substituted by an exceptional divisor coincident with $k-1$
intersecting copies of $\mathbb{P}^1$, in such a way
that the neighborhood of the point becomes an ALE space ${\cal{O}}_{\mathbb{P}^1} (-k)$, itself locally isomorphic to $\mathbb{C}^2/\mathbb{Z}_k$. The
resolved space obtained, for $k>3$, by successive resolutions of all the singularities of different orders, exhibits non-trivial cycles whose intersection matrix is the opposite of the Cartan matrix of the corresponding $\mathbb{A}_{k-1}$ algebra\footnote{We do not consider more general (non-abelian) orbifolds where the intersection matrix is related to algebras of D and E type.}.  It is also easy to verify that, after the blow-up, the resulting manifolds have $\chi=24$.

Finally, let us  discuss the integer homology $H^2 (K3,\mathbb{Z})$ of the orbifold, starting from the description of the integer homology of the covering torus,  contained in the Hodge diamond
\be
\begin{array}{ccccc} &&h_{00}\\ &h_{10}&&h_{01}\\ h_{20}&&h_{11}&&h_{02}\\
&h_{21}&&h_{12} \\ &&h_{22} \end{array}=
\begin{array}{ccccc} &&1\\ &2&&2\\ 1&&4&&1\\
&2&&2 \\ &&1 \end{array} \ ,\label{torushodgediamond}
\ee
where, of course, $h_{pq}=dim_{\mathbb{C}} H^{pq}(\mathbb{C})$. The non-trivial part is represented by the two-forms corresponding to
\be
\sigma_{ij} = dx^i \wedge dx^j \ ,
\ee
with $\{x^i\}$ the real coordinates of the torus.  It is useful to introduce the basis ($\epsilon_{ijkl}$ is the Levi-Civita symbol)
\be
\lambda^{\pm}_{ij} = \sigma_{ij} \, \pm \frac{1}{2} \, \epsilon_{ijkl} \, \sigma_{kl} =  \sigma_{ij} \, \pm \tilde{\sigma}_{ij}\ ,
\ee
that makes manifest the properties of the forms under Hodge duality. Looking at the corresponding expressions in
complex coordinates,
\begin{align}
\lambda^{\pm}_{12}&=\Pi_{1 \bar{1}} \, \pm \, \Pi_{2 \bar{2}} \ , \nonumber \\
\lambda^{+}_{13}&=\Pi_{12} \, + \, \Pi_{\bar{1} \bar{2}} \ , \nonumber \\
\lambda^{-}_{13}&=\Pi_{1 \bar{2}} \, + \, \Pi_{\bar{1} 2} \ ,\nonumber \\
\lambda^{+}_{14}&=\Pi_{12} \, - \, \Pi_{\bar{1} \bar{2}} \ ,\nonumber \\
\lambda^{-}_{14}&=\Pi_{1 \bar{2}} \, - \, \Pi_{\bar{1} 2} \ ,
\end{align}
where $\Pi_{\alpha \beta}=dz_{\alpha} \wedge dz_{\beta}$, it can be observed that
three out of four combinations of the $(1,1)$-forms are anti-self-dual, while $\lambda^{+}_{13}$ and $\lambda^{+}_{14}$, combinations of the
holomorphic $(2,0)$-form and its complex-conjugate $(0,2)$-form, are self-dual.  It is important to notice that only
the self-dual forms, together with $\lambda^-_{12}$, are well defined on the corresponding $\mathbb{Z}_N$-orbifold, being invariant
under the action in (\ref{orbaction}).  The $\mathbb{Z}_2$-orbifold is an exception, since in that case all the two-forms are invariant.
Calculating the scalar product results in
\be
(\sigma_{ij}, \sigma_{kl}) = \epsilon_{ijkl} \ ,
\ee
once the proper normalization to 1 of the volume of the torus is chosen.  It gives rise to the intersection matrix
\be
\Gamma^{3,3} = U(-1)\oplus U(-1)\oplus U(-1)
\ee
for the homology cycles dual to the sigmas, where
\be
U(-k)= \left(\begin{array}{cc}0 & k \\ k & 0 \end{array}\right) \ .
\ee

On the orbifolds, as previously mentioned, some of the cycles are inherited by the torus ones.
Indeed, a ``bulk cycle'' is built by considering the group invariant combination of the torus cycles
\be
\pi^{ij}= \frac{1}{N} \ \bigoplus_{g\in\mathbb{Z}_N} g\,  \sigma^{ij} \ .
\ee
As checked before, only four cycles survive the projection, with the exception of the $\mathbb{Z}_2$ case where, accidentally, all the cycles are
invariant and can be promoted to bulk cycles of the orbifold.  Moreover, the integer homology must be completed adding the exceptional divisors, whose
volume is zero in the metric inherited from the torus, resulting in a vanishing intersection number with the bulk cycles.  Their mutual intersection, as seen before, is described by (minus) the Cartan matrix of the corresponding lattice.  The related Hodge diamond, after the orbifold quotient, becomes
\be
\begin{array}{ccccc} &&h_{00}\\ &h_{10}&&h_{01}\\ h_{20}&&h_{11}&&h_{02}\\
&h_{21}&&h_{12} \\ &&h_{22} \end{array}=
\begin{array}{ccccc} &&1\\ &0&&0\\ 1&&1^+ + 19^-&&1\\
&0&&0\\ &&1 \end{array} \ ,
\ee
where the corresponding self-duality properties have been stressed.

It can be recognized that,
after the blow-up, the orbifold produces a K3 manifold.  Indeed, in the $\mathbb{Z}_N$ case with $N=3,4,6$ one has 4 invariant forms
and $18$ exceptional divisors, while in the $N=2$ case there are $6$ invariant forms and $16$ exceptional divisors.
The corresponding smooth manifolds
have Euler characteristic $24$ and Hirzebruch signature $16$, as expected.  Moreover, the homological two-cycles reflect the
structure of the dual forms and give rise to a lattice with the signature $(3,19)$.  However, the construction of an
homology basis for the $\Gamma^{(3,19)}$ lattice is more subtle than simply the superposition of bulk cycles and exceptional divisors.
To clarify this central point let's analyze, for simplicity, the $\mathbb{Z}_2$ model,
where the exceptional divisors in the twisted directions are all of order two, with intersections given by $(E_i, E_j) = -2 \delta_{ij}, i,j=1,...,16$.  The  intersection matrix, including the bulk cycles, takes then the form
\be
\Sigma=U(-2)\oplus U(-2)\oplus U(-2)\oplus (-2) \ \mathbb{I}_{16} \ ,
\ee
where $\mathbb{I}_d$ is the $d$-dimensional identity matrix. It is evident that the basis we have considered cannot be the right one for the integer homology, since the related lattice must be \cite{Aspinwall:1994rg} an even and self-dual lattice of signature $(3,19)$, while, for instance, $det(\Sigma)= - 2^{22}$.  A correct basis can be obtained as follows.  The key observation is that the bulk cycles make sense only if their are chosen in a generic position.  Indeed, if a cycle homologous to a bulk cycle intersect some of the fixed points, the combination of the bulk cycle with the exceptional cycles related to the fixed points it meets is a double (or multiple) cover of the K3 cycle \cite{Wendland:2000ry, Nahm:1999ps}.  The related objects are very familiar in string theory: they correspond to the so-called ``fractional branes'' \cite{Polchinski:1998rq, Blumenhagen:2002wn}.  They cannot be moved away from the fixed points, unless they are combined with the (Ramond-Ramond) antibranes to give rise to a bound state that can be identified with a ``bulk'' brane.  Moreover, differences of equivalent fractional branes must belong, for consistency, to the homology of the orbifold.  As a result, $H^2(K3,\mathbb{Z})$ comprises the ``fractional cycles'' and the fraction of exceptional divisors that ``intertwine'' among different fractional branes.

Explicit constructions of such kind of basis for the ${\mathbb{Z}}_N$-orbifolds can be found in \cite{Wendland:2000ry}.  The matrix corresponding to the intersection form is, in general, quite complicated.  However, since the even self-dual lattice are quite
rare, it is always possible to choose a basis in which it takes the form \cite{Serre}
\be
\Gamma^{3,19}=U(-1)\oplus U(-1)\oplus U(-1)\oplus \Gamma_8(-1) \oplus \Gamma_8(-1) \ ,
\ee
where $\Gamma_8(-1)$ is the opposite of the Cartan matrix of $e_8$.  As is well known, the $\Gamma^{3,19}$ parameterize the charges of the BPS states obtained from D-branes wrapped on cycles of K3.  The crucial point is that, whenever the right basis is chosen, there exists always a ``light-like'' sublattice
\be
\Gamma^{3,3} \ \subset \Gamma^{3,19} \ .
\ee
Along the relative six directions, corresponding to six two-cycles, the D-branes  wrap to give single $1/2$-BPS states. These wrapped D2-branes, together with the unwrapped D0-branes and the D4-branes wrapped on the whole of K3, are identified under the Heterotic-Type IIA duality with
the perturbative Heterotic states described in Section $2$.

This refines the conjecture in ref. \cite{Bergshoeff:2012jb} where the six cycles on which the wrapping rules are applied were identified with the bulk cycles. For any orbifold one can consider the wrapping rules of the branes wrapping the fundamental 2-cycles  corresponding to the light-like directions in the $\Gamma^{3,19}$ lattice, and we conjecture that this generalizes to any smooth K3 manifold. In the following, we shall show that this conjecture gives the right number of branes in the IIA theory on the orbifold as predicted by the Heterotic-Type IIA duality. First, in the next subsection, we shall review how T-duality relates the IIA theory on a K3 orbifold to the IIB theory on a non-geometric orbifold. This leads to another check of the wrapping rules.

\subsection{T-duality and non-geometric orbifolds of Type IIB}

T-duality \cite{Giveon:1994fu} is a perturbative transformation of the full quantum string theory.  The simplest example can be obtained by considering Type IIA string compactified on a circle of radius $R$.  T-duality, in this case, acts as a right-moving target-space parity on the string bosonic and fermionic coordinates,  namely $(X^9_L, \psi^9_L) \to (X^9_L, \psi^9_L)$ and $(X^9_R, \psi^9_R) \to (-X_R, -\psi^9_R)$. The result is a mapping of parameters and a flip of the chirality of the Ramond ground state, in such a way that the Type IIA string compactified on a circle of radius $R$ results into a theory that is equivalent to a Type IIB string compactified on a circle of radius $\alpha'/R$. In particular, the even R-R potential forms are mapped into the odd R-R potential forms of Type IIB, signaling the fact that the corresponding D$p$-branes become D${(p-1)}$-branes (D${(p+1)}$-branes) if the direction $9$ is longitudinal (transverse) to the D-brane world-volume.
An even number of such transformations is a symmetry of Type IIA and Type IIB strings, while an odd number of them is a duality among the two.
In more general toroidal compactifications, the structure of the T-duality group remains the same. In particular, since the D-branes are in a spinorial representation of the group, those transformations that flip the chirality of the spinor representation of $\text{SO}(d,d)$ are T-dualities that map Type IIA into Type IIB, while the remaining transformations are global symmetries of the Type IIA or Type IIB strings.

On K3, T-duality is difficult to understand in general, but again one can analyze what happens in the orbifold limits.  The {\it geometric} orbifold action on the $z_1$ coordinate of eq. (\ref{orbaction}) is the same on the left and right components of the corresponding string coordinates, namely
\be
z_1^L \rightarrow g \, z_1^L \ , \quad z_1^R \rightarrow g \, z_1^R \ ,
\ee
where, if $z_1$ is identified with the two-torus coordinate in the directions $8$ and $9$, $z_1^L=X_8^L + i X_9^L$ and $z_1^R=X_8^R + i X_9^R$.  On the other hand, a T-duality along $X_9$ combined with the orbifold action provides
\be
z_1^L \rightarrow g \, z_1^L \ , \quad z_1^R \rightarrow g \, \bar{z}_1^R \ ,
\ee
accompanied by the corresponding exchange of Dirichlet with Neumann boundary conditions for the D-branes along the ninth direction.
In other words, the Type IIA compactified on an orbifold limit of K3 is mapped onto a non-geometric compactification of the Type IIB.

In general, the wrapping rules capture the nature of the geometry that a specific brane probes, and branes with different $\alpha$ probe different geometries. In principle, this should be the case regardless of whether the `true' geometry, {\it i.e.} the geometry that a point particle probes, is actually Riemannian or not. We are therefore tempted to consider the non-geometric nature of the Type IIB orbifold as leading to a non-geometric way of counting the cycles along which the D-branes of Type IIB theory wrap, leaving the wrapping rules unchanged.
The D-brane behaviour in Orientifolds teaches us the way to proceed.  It is well known that T-duality exchanges generically the $B$-field and (geometric) components of the metric \cite{Angelantonj:1999xf, Pradisi:1999ii}.  Inside the {\it quantum} homology of K3, this exchange can be identified with an internal automorphism that redefines the basis of the lattice $\Gamma^{4,20}$ along the light-cone directions.  Let us apply this observation to the $1/2$-BPS D0-branes of the Type IIA.  They come from the ten dimensional unwrapped D0 brane, the D2 branes wrapping the six two cycles of $\Gamma^{3,3}$ and the D4-brane wrapping the whole K3.  A $X_9$ T-duality will reflect itself into a T-duality along one of the $\Gamma^{4,4}$ light-cone directions. As a result, it is easy to verify that the unwrapped D0 and the 3 D2 wrapped longitudinally to the T-dualized direction are mapped to the D1 branes of the Type IIB wrapping a one-cycle, while the remaining four are mapped to D3 branes of the Type IIB wrapping a three-cycle.  In other words, there is a mixing of the geometric form potentials with the B-field that gives rise to a T-dual basis of forms, whose dual cycles can be identified with the four one-cycles in $\Gamma^{2,2}=U(-1) \oplus U(-1)$ and the four three-cycles in the other $\Gamma^{2,2}=U(-1) \oplus U(-1)$, forming together a (new) basis of $\Gamma^{4,4}=\Gamma^{2,2}\oplus \ \Gamma^{2,2}$.  Clearly, in the Type IIB non-geometric orbifold compactification, a choice of basis that can be identified with the original (purely geometrical) basis of K3 is absent.

We now consider how the other branes in the two theories see the respective geometries.
The wrapped fundamental string  and the pp-wave are interchanged  as the corresponding momentum and winding modes under T-duality, while the unwrapped fundamental string clearly sees the same geometry in the two theories. This implies that the $\alpha=0$ branes see the standard K3 geometry in both cases. Electromagnetic duality relates the branes with $\alpha=-2$ to the ones with $\alpha=0$ and the branes with $\alpha=-3$ to the ones with $\alpha=-1$.
We are therefore led to conjecture that our way of counting cycles in the non-geometric orbifold
extends to all the non-perturbative solitons as follows: the  branes with odd $\alpha$ wrap odd cycles because they are sensible to the T-dual direction, while the branes with even $\alpha$ wrap even cycles because they are blind to it.   In particular, branes absent in Type IIA but present in the non-geometric Type IIB orbifold account for the corresponding dual states in the Heterotic string on $T^4$.  In the next section, we shall show in more detail how these rules exactly match the $SO(4,4)$ triality that, as already observed \cite{Kiritsis:2000zi, Nahm:1999ps}, is at the heart of the Heterotic-Type IIA-Type IIB string-string dualities in $D=6$.  Moreover, a generalized geometry structure seems to  emerge quite clearly where the wrapping rules, as usual, take care of the non-standard way in which the exotic branes see the geometry.  It would be very interesting to verify directly these rules on the corresponding classical supergravity solutions.

\subsection{Non-perturbative branes and wrapping rules}

Our analysis so far was based on mapping the fundamental 0-branes of the Heterotic string to the D0-branes of the Type-II theories on K3. For the case of the Type IIA D-branes, we have shown in subsection 3.2 that one has to consider the wrapping rules on a ``light-like'' basis of homology 2-cycles, while in subsection 3.3 we have shown that, by T-duality, the same theory corresponds to a non-geometric orbifold of the Type IIB. Within the $T^4/\mathbb{Z}_N$ orbifold of the Type IIA, one can count the D-branes in 6 dimensions as wrapping six fundamental two-cycles of the orbifold. Similarly, from the T-dual Type IIB perspective, one should  formally consider the bulk 1-cycles and 3-cycles to get the right counting. The number of such cycles is 4, as explained in the previous section, and the wrapping of D1-branes on four 1-cycles and of D3-branes on four 3-cycles indeed results into eight D0-branes.
Let us show how this extends to the other branes of the Type-II strings. As previously mentioned, denoting with $\alpha_{Het}$  and $\alpha_{II}$ the dilaton scaling of the tension of a $p$-brane in the Heterotic and in the Type-II theories, the duality relates them according to the eq. \eqref{alphahetalphaIIA}.  It allows us to check the wrapping rules in both Type IIA and Type IIB strings for all the branes.  In particular, the wrapping rules in the Type IIB case have to be generalized as follows: the branes with even $\alpha$ wrap on even cycles while the branes with odd $\alpha$ wrap on odd cycles.

Considering the K3 orbifold limits, it is natural to write the charges as representations of  $\text{SO}(4,4)$. From this point of view, the map relating the Heterotic string on $T^4$, the Type IIA string on the geometric orbifold and the Type IIB string on the non-geometric one, corresponds to the triality relating vector, spinor and conjugate-spinor representations of $\text{SO}(4,4)$, as summarized in Table \ref{SO44triality}. For the 0-branes, only the first row in Table \ref{SO44triality} is relevant, because the fundamental 0-brane BPS states of the Heterotic theory are in the ${\bf 8}_{\rm  V}$, while the wrapping of even D-branes on even cycles gives the ${\bf 8}_{\rm  S}$ of Type IIA, and the wrapping of odd D-branes on odd cycles gives the ${\bf 8}_{\rm  C}$ of Type IIB.

\begin{table}[h]
\begin{center}
\begin{tabular}{|c|c|c|}
\hline \rule[-1mm]{0mm}{6mm} Het & IIA & IIB \\
\hline \hline \rule[-1mm]{0mm}{6mm} ${\bf 8}_{\rm V}$ &  ${\bf 8}_{\rm S}$  &  ${\bf 8}_{\rm C}$ \\
\hline \rule[-1mm]{0mm}{6mm} ${\bf 8}_{\rm S}$ &  ${\bf 8}_{\rm C}$  &  ${\bf 8}_{\rm V}$\\
\hline \rule[-1mm]{0mm}{6mm} ${\bf 8}_{\rm C}$ &  ${\bf 8}_{\rm V}$  &  ${\bf 8}_{\rm S}$\\
\hline
\end{tabular}
\caption{\sl \footnotesize The triality mapping the  $\text{SO}(4,4)$ representations of the Heterotic on $T^4$, the IIA on $T^4/\mathbb{Z}_N$ and the IIB on the $\mathbb{Z}_N$ non-geometric orbifold.
\label{SO44triality}}
\end{center}
\end{table}

Let us now use Table \ref{SO44triality} in conjunction with eq. \eqref{alphahetalphaIIA} to find the mapping between the other brane charges
of the three string theories.
The fundamental Heterotic string is mapped to a solitonic $\alpha_{II}=-2$ string on the IIA side. The latter is the NS5-brane wrapped on K3 that, according to the wrapping rule, does not double.  As Table \ref{branesD=6heterotic} shows,
the Heterotic theory also contains branes with $\alpha_{Het}=-2$ and $\alpha_{Het}=-4$.
The branes with  $\alpha_{Het}=-2$ correspond to branes with $\alpha_{II} = -p +1$ on the IIA side, so to branes with $\alpha_{II}=0,-1,-2$ for $p=1, 2, 3$, respectively.  They can be derived from the ten dimensional Type IIA string, where they are already present.  Indeed, for $p=1$, one gets the fundamental string of the Type IIA.  For $p=2$, one gets a charge $Q_A$ which is mapped to an $\alpha_{II}=-1$ 2-brane with charge $Q_a$ in Type IIA and $Q_{\dot{a}}$ in Type IIB, where $a$ and $\dot{a}$ are the two spinor indices of  $\text{SO}(4,4)$. Using the wrapping rules for the Type IIA D-branes on even cycles and the Type IIB D-branes on odd cycles one gets the right numbers. The charges of the 3-branes are fixed under triality. For $\alpha_{Het} =-2$ and $p=4$, one gets instead a 4-brane of the Heterotic theory with charge $Q_{ABC}$, with three antisymmetric indices of $\text{SO}(4,4)$, which is mapped to an $\alpha=-3$ 4-brane with  charge  $Q_{A\dot{a}}$ in Type IIA and $Q_{Aa}$ in Type IIB. The branes associated to the latter charge can be obtained, following the last rule in eq. \eqref{allwrappingrulesinonego}, as the $\alpha=-3$ 7-brane of Type IIB wrapped on a 3-cycle.  They amount to a total of 32 4-branes, exactly as expected from the Heterotic side.
Finally, for $\alpha_{Het} =-2$ and $p=5$, one gets the two irreducible representations $Q_{ABCD}^+$ and $Q_{ABCD}^-$, that are self-dual and anti-self-dual, respectively. The former is mapped to an $\alpha_{II}=-4$ charge $Q_{AB}$ (with two symmetric indices) of the Type IIA or to an $\alpha_{II}=-4$ charge $Q_{ABCD}^-$ in Type IIB. The latter is mapped to a charge $Q_{ABCD}^-$ in Type IIA and to $Q_{AB}$ (with two symmetric indices) in Type IIB.
The branes with charges $Q_{ABCD}^-$ of Type IIB correspond exactly to the ten dimensional $\alpha=-4$ branes,  if one uses the wrapping rules in eq. \eqref{alpha=-4wrappingrules}.

\begin{table}[h!]
\begin{center}
\begin{tabular}{|c||c|c|c|c|c|c|}
\hline \rule[-1mm]{0mm}{6mm} \backslashbox{$\alpha$}{$p$} & $p=0$ & $p=1$ & $p=2$ & $p=3$ & $p=4$ &$p=5$ \\
\hline \hline \rule[-1mm]{0mm}{6mm} $\alpha=0$ &  & 1& & & & \\
\hline \rule[-1mm]{0mm}{6mm} $\alpha=-1$ & 8 & & 8 && 8&  \\
\hline \rule[-1mm]{0mm}{6mm} $\alpha=-2$ & & 1&&24 & &8\\
\hline
\end{tabular}
\caption{\sl \footnotesize The number of 1/2-BPS branes  of the Type IIA theory compactified on an orbifold of K3 that can be obtained applying the wrapping rules of eq. \eqref{allwrappingrulesinonego} to the 10-dimensional Type IIA branes for the different values of $\alpha$.
\label{branesD=6IIA}}
\end{center}
\end{table}

Let us now consider the branes with $\alpha_{Het} =-4$. For $p=4$, the charge $Q_A$ in the Heterotic string is mapped, by triality, to a brane with $\alpha_{II}=-1$ and charge $Q_a$ in Type IIA and $\alpha_{II}=-1$ and charge $Q_{\dot{a}}$ in Type IIB. They can also be obtained using wrapping rules from D4, D6 and D8-branes of Type IIA on even cycles and from D5, D7 and  D9-branes of Type IIB on odd cycles.  Finally, for $p=5$ the charge $Q_{AB}$  (with two symmetric indices) on the Heterotic side is mapped to an $\alpha_{II}=-2$ brane with charge $Q_{ABCD}^-$ in Type IIA and $Q_{ABCD}^+$ in Type IIB. In both cases, they correspond to the unwrapped NS5-brane. The third wrapping rule in eq. \eqref{allwrappingrulesinonego} would then give a total of 16 branes, but the number  must actually be halved because of the induced self-duality condition on the charge.

\begin{table}[h!]
\begin{center}
\begin{tabular}{|c||c|c|c|c|c|c|}
\hline \rule[-1mm]{0mm}{6mm} \backslashbox{$\alpha$}{$p$} & $p=0$ & $p=1$ & $p=2$ & $p=3$ & $p=4$ &$p=5$ \\
\hline \hline \rule[-1mm]{0mm}{6mm} $\alpha=0$ &  & 1& & & & \\
\hline \rule[-1mm]{0mm}{6mm} $\alpha=-1$ & 8 & & 8 && 8&  \\
\hline \rule[-1mm]{0mm}{6mm} $\alpha=-2$ & & 1&&24 & &8\\
\hline \rule[-1mm]{0mm}{6mm} $\alpha=-3$ & & && & 32&\\
\hline \rule[-1mm]{0mm}{6mm} $\alpha=-4$ & & && & &8\\
\hline
\end{tabular}
\caption{\sl \footnotesize  The number of 1/2-BPS branes  of the Type IIB theory compactified on a non-geometric orbifold of K3  that can be obtained applying the wrapping rules of eq. \eqref{allwrappingrulesinonego} to the 10-dimensional Type IIB branes for the different values of $\alpha$, where the branes wrap even cycles for even $\alpha$ and odd cycles for odd $\alpha$.
\label{branesD=6hIIB}}
\end{center}
\end{table}

The outcome of this whole analysis is that, if one considers both Type IIA and Type IIB theories, only the 8 Heterotic $\alpha_{Het}=-2$ branes with charge $Q_{ABCD}^-$ cannot be seen, on the Type-II side, as coming from ten-dimensional wrapped branes.  This is completely consistent, because it is not expected that branes satisfying wrapping rules in one theory necessarily satisfy dual wrapping rules in the dual theory.  Indeed, the same phenomenon occurs for the $\alpha=-4$ branes on the Heterotic side dual to $\alpha=-1,-2$-branes on the Type II side. To summarize, in Tables \ref{branesD=6IIA} and \ref{branesD=6hIIB} we report all the branes of the 6-dimensional Type II theory, obtained using the wrapping rules from the Type IIA and Type IIB theories in $D=10$.

\section{Conclusions}

In this paper we exploited the Heterotic-Type II string-string duality in $D=6$ to derive the wrapping rules
obeyed by Type IIA branes in K3 compactifications.  This duality relates perturbative 1/2-BPS 0-branes on the Heterotic side
to D0 branes on the Type IIA side.  The latter correspond, from a ten dimensional perspective, to unwrapped D0 branes, D4 branes wrapping the whole K3 and D2 branes wrapping homology two-cycles. In the $T^4/\mathbb{Z}_N$ orbifold limits of K3, where a Conformal Field Theory description is available, we were able to identify a basis of fundamental homology cycles on which the wrapping rules correctly account for the single brane states, in agreement with duality.
We then applied the wrapping rules to other solitonic branes of the Type IIA, finding agreement with the Heterotic dual 1/2-BPS states.
Finally, we used T-duality, mapping geometric $T^4/\mathbb{Z}_N$ orbifolds of Type IIA  to non-geometric orbifolds of Type IIB, in order to
further extend the wrapping rules to the branes of the ten-dimensional Type IIB string.  In that respect, a crucial role is played by the SO(4,4) triality relating, in this context, Heterotic, Type IIA and Type IIB theories.

These results are a generalization to the $T^4/\mathbb{Z}_N$ orbifold of those obtained in \cite{Bergshoeff:2012jb} for the case of $T^4/\mathbb{Z}_2$. In the same paper, also the $T^4/\mathbb{Z}_2$ orbifold of the Type IIB theory was considered, showing again that the wrapping rules exactly reproduce the spectrum of 1/2-BPS single branes of the six-dimensional ${\cal N}=(2,0)$ theory.  It should be noticed that the whole analysis can be straightforwardly generalized to any $T^4/\mathbb{Z}_N$ orbifold of the Type IIB, following exactly the same construction that was discussed in this paper.

It would be very interesting to further extend this analysis to wider classes of compactification manifolds,
like for instance Calabi-Yau threefolds, using additional duality relations in lower dimensions and to better understand
the role of (generalized) fractional branes.
Another very interesting direction would be to elucidate the connection between the wrapping rules and
the  doubled geometry approach discussed in
\cite{Hull:2004in}. In particular, our way to extend the wrapping rules to curved internal spaces could suggest
a direction to go beyond toroidal compactification in the doubled-geometry context.

On more general grounds, one could also conceive the possibility of connecting our discussion to
K-Theory, related not only to the (co)homology of the compactification manifolds but also to the gauge-bundle structure of the
space of R-R charges.  Perhaps, there is a generalization of K-Theory to a more
complicated structure that is able to describe not only the R-R charges of the D-branes but also
the charges of the other BPS branes.

To conclude, trying to uncover the underlying structure of String/M-Theory leads one to deal with a plethora of
extended objects, that are beyond the reach of perturbation theory and have increasing negative power-like dependence on the
string coupling.  It would be of great importance to better understand the role they can play in the many ramifications and applications of the theory like, for instance, in the AdS/CFT correspondence and in the Black Hole Physics \cite{deBoer:2012ma}.

\section{Acknowledgments}

It is a pleasure to thank E. Dudas and A. Sagnotti for very interesting discussions.  G.P. would like to thank K. Wendland for e-mail correspondence.
G.P. and C.C. would like to thank the CPHT, Ecole Polytechnique,  for the kind hospitality while this work was being completed.
F.R. would like to thank the University of Groningen for hospitality at various stages of this work.
E.B. would like to thank the Universit\`a di Roma ``La Sapienza''  for its hospitality.
This work was partially supported by the Italian MIUR-PRIN contract 2009-KHZKRX.


\begin{thebibliography}{99}

\bibitem{Witten:1978mh}
  E.~Witten and D.~I.~Olive,
  ``Supersymmetry Algebras That Include Topological Charges,''
  Phys.\ Lett.\ B {\bf 78} (1978) 97.

\bibitem{Polchinski:1998rq}
  See, {\it e.g.}, J.~Polchinski,
  ``String theory. Vol. 1: An introduction to the bosonic string,''
  Cambridge, UK: Univ. Pr. (1998) 402 p ;
  ``String theory. Vol. 2: Superstring theory and beyond,''
  Cambridge, UK: Univ. Pr. (1998) 531 p. ;
   C.~V.~Johnson,
  ``D-branes,''
  Cambridge, USA: Univ. Pr. (2003) 548 p.

\bibitem{Angelantonj:2002ct}
  See, {\it e.g.}, C.~Angelantonj and A.~Sagnotti,
  ``Open strings,''
  Phys.\ Rept.\  {\bf 371} (2002) 1
   [Erratum-ibid.\  {\bf 376} (2003) 339]
  [hep-th/0204089].

\bibitem{Townsend:1996xj}
  See, {\it e.g.}, P.~K.~Townsend,
  ``P-brane democracy,''
  In *Duff, M.J. (ed.): The world in eleven dimensions* 375-389
  [hep-th/9507048];
  P.~K.~Townsend,
  ``Four lectures on M theory,''
  In *Trieste 1996, High energy physics and cosmology* 385-438
  [hep-th/9612121].

\bibitem{Hull:1994ys}
  C.~M.~Hull and P.~K.~Townsend,
  ``Unity of superstring dualities,''
  Nucl.\ Phys.\ B {\bf 438} (1995) 109
  [hep-th/9410167].

\bibitem{Bergshoeff:2011qk}
  E.~A.~Bergshoeff and F.~Riccioni,
  ``The D-brane U-scan,''
  arXiv:1109.1725 [hep-th].

\bibitem{Kleinschmidt:2011vu}
  A.~Kleinschmidt,
  ``Counting supersymmetric branes,''
  JHEP {\bf 1110} (2011) 144
  [arXiv:1109.2025 [hep-th]].

\bibitem{Bergshoeff:2012ex}
  E.~A.~Bergshoeff, A.~Marrani and F.~Riccioni,
  ``Brane orbits,''
  Nucl.\ Phys.\ B {\bf 861} (2012) 104
  [arXiv:1201.5819 [hep-th]].

\bibitem{Bergshoeff:2011mh}
  E.~A.~Bergshoeff and F.~Riccioni,
  ``Dual doubled geometry,''
  Phys.\ Lett.\ B {\bf 702} (2011) 281
  [arXiv:1106.0212 [hep-th]].

\bibitem{Bergshoeff:2005ac}
  E.~A.~Bergshoeff, M.~de Roo, S.~F.~Kerstan and F.~Riccioni,
  ``IIB supergravity revisited,''
  JHEP {\bf 0508} (2005) 098
  [hep-th/0506013];
    E.~A.~Bergshoeff, J.~Hartong, P.~S.~Howe, T.~Ortin and F.~Riccioni,
  ``IIA/IIB Supergravity and Ten-forms,''
  JHEP {\bf 1005} (2010) 061
  [arXiv:1004.1348 [hep-th]].

\bibitem{Bergshoeff:2006qw}
  E.~A.~Bergshoeff, M.~de Roo, S.~F.~Kerstan, T.~Ortin and F.~Riccioni,
  ``IIA ten-forms and the gauge algebras of maximal supergravity theories,''
  JHEP {\bf 0607} (2006) 018
  [hep-th/0602280].


\bibitem{Riccioni:2007au}
  F.~Riccioni and P.~C.~West,
  ``The E(11) origin of all maximal supergravities,''
  JHEP {\bf 0707} (2007) 063
  [arXiv:0705.0752 [hep-th]].

\bibitem{Bergshoeff:2007qi}
  E.~A.~Bergshoeff, I.~De Baetselier and T.~A.~Nutma,
  ``E(11) and the embedding tensor,''
  JHEP {\bf 0709} (2007) 047
  [arXiv:0705.1304 [hep-th]].

\bibitem{Bergshoeff:2011zk}
  E.~A.~Bergshoeff and F.~Riccioni,
  ``String Solitons and T-duality,''
  JHEP {\bf 1105} (2011) 131
  [arXiv:1102.0934 [hep-th]].

\bibitem{Bergshoeff:2011ee}
  E.~A.~Bergshoeff and F.~Riccioni,
  ``Branes and wrapping rules,''
  Phys.\ Lett.\ B {\bf 704} (2011) 367
  [arXiv:1108.5067 [hep-th]].

\bibitem{Bergshoeff:2012jb}
  E.~A.~Bergshoeff and F.~Riccioni,
  ``Heterotic wrapping rules,''
  JHEP {\bf 1301} (2013) 005
  [arXiv:1210.1422 [hep-th]].


\bibitem{Witten:1995ex}
  E.~Witten,
  ``String theory dynamics in various dimensions,''
  Nucl.\ Phys.\ B {\bf 443} (1995) 85
  [hep-th/9503124].

\bibitem{Aspinwall:1995zi}
  P.~S.~Aspinwall,
  ``Enhanced gauge symmetries and K3 surfaces,''
  Phys.\ Lett.\ B {\bf 357} (1995) 329
  [hep-th/9507012].

\bibitem{Bergshoeff:2013sxa}
  E.~A.~Bergshoeff, F.~Riccioni and L.~Romano,
  ``Branes, Weights and Central Charges,''
  JHEP {\bf 1306} (2013) 019
  [arXiv:1303.0221 [hep-th]].

\bibitem{Greene:1989ya}
  B.~R.~Greene, A.~D.~Shapere, C.~Vafa and S.~-T.~Yau,
  ``Stringy Cosmic Strings and Noncompact Calabi-Yau Manifolds,''
  Nucl.\ Phys.\ B {\bf 337} (1990) 1.

\bibitem{Bergshoeff:2006jj}
  E.~A.~Bergshoeff, J.~Hartong, T.~Ortin and D.~Roest,
  ``Seven-branes and Supersymmetry,''
  JHEP {\bf 0702} (2007) 003
  [hep-th/0612072].

\bibitem{Dabholkar:1995nc}
  A.~Dabholkar, J.~P.~Gauntlett, J.~A.~Harvey and D.~Waldram,
  ``Strings as solitons and black holes as strings,''
  Nucl.\ Phys.\ B {\bf 474} (1996) 85
  [hep-th/9511053].

\bibitem{Schwarz:1993mg}
  J.~H.~Schwarz and A.~Sen,
  ``Duality symmetries of 4-D heterotic strings,''
  Phys.\ Lett.\ B {\bf 312} (1993) 105
  [hep-th/9305185].


\bibitem{Dabholkar:1990yf}
  A.~Dabholkar, G.~W.~Gibbons, J.~A.~Harvey and F.~Ruiz Ruiz,
  ``Superstrings and Solitons,''
  Nucl.\ Phys.\ B {\bf 340} (1990) 33.


\bibitem{Callan:1995hn}
  C.~G.~Callan, J.~M.~Maldacena and A.~W.~Peet,
  ``Extremal black holes as fundamental strings,''
  Nucl.\ Phys.\ B {\bf 475} (1996) 645
  [hep-th/9510134].


\bibitem{Narain:1985jj}
  K.~S.~Narain,
  ``New Heterotic String Theories in Uncompactified Dimensions $<$ 10,''
  Phys.\ Lett.\ B {\bf 169} (1986) 41.

\bibitem{Narain:1986am}
  K.~S.~Narain, M.~H.~Sarmadi and E.~Witten,
  ``A Note on Toroidal Compactification of Heterotic String Theory,''
  Nucl.\ Phys.\ B {\bf 279} (1987) 369.

\bibitem{Giveon:1994fu}
  For a review see, {\it e.g.}, A.~Giveon, M.~Porrati and E.~Rabinovici,
  ``Target space duality in string theory,''
  Phys.\ Rept.\  {\bf 244} (1994) 77
  [hep-th/9401139].

\bibitem{Aspinwall:1994rg}
  P.~S.~Aspinwall and D.~R.~Morrison,
  ``String theory on K3 surfaces,''
  In *Greene, B. (ed.), Yau, S.T. (ed.): Mirror symmetry II* 703-716
  [hep-th/9404151];
P.~S.~Aspinwall,
  ``K3 surfaces and string duality,''
  In *Yau, S.T. (ed.): Differential geometry inspired by string theory* 1-95
  [hep-th/9611137].

\bibitem{Nahm:1999ps}
  W.~Nahm and K.~Wendland,
  ``A Hiker's guide to K3: Aspects of N=(4,4) superconformal field theory with central charge c = 6,''
  Commun.\ Math.\ Phys.\  {\bf 216} (2001) 85
  [hep-th/9912067].


\bibitem{Seiberg:1988pf}
  N.~Seiberg,
  ``Observations on the Moduli Space of Superconformal Field Theories,''
  Nucl.\ Phys.\ B {\bf 303} (1988) 286.



\bibitem{Sen:1995cj}
  A.~Sen,
  ``String string duality conjecture in six-dimensions and charged solitonic strings,''
  Nucl.\ Phys.\ B {\bf 450} (1995) 103
  [hep-th/9504027].

\bibitem{Harvey:1995rn}
  J.~A.~Harvey and A.~Strominger,
  ``The heterotic string is a soliton,''
  Nucl.\ Phys.\ B {\bf 449} (1995) 535
   [Erratum-ibid.\ B {\bf 458} (1996) 456]
  [hep-th/9504047].

\bibitem{Strominger:1995cz}
  A.~Strominger,
  ``Massless black holes and conifolds in string theory,''
  Nucl.\ Phys.\ B {\bf 451} (1995) 96
  [hep-th/9504090].


\bibitem{Kiritsis:2000zi}
  E.~Kiritsis, N.~A.~Obers and B.~Pioline,
  ``Heterotic / type II triality and instantons on K(3),''
  JHEP {\bf 0001} (2000) 029
  [hep-th/0001083].

\bibitem{Ooguri:1996ck}
  H.~Ooguri, Y.~Oz and Z.~Yin,
  ``D-branes on Calabi-Yau spaces and their mirrors,''
  Nucl.\ Phys.\ B {\bf 477} (1996) 407
  [hep-th/9606112].

\bibitem{Bershadsky:1995qy}
  M.~Bershadsky, C.~Vafa and V.~Sadov,
  ``D-branes and topological field theories,''
  Nucl.\ Phys.\ B {\bf 463} (1996) 420
  [hep-th/9511222].


\bibitem{Callan:1987px}
 C.~G.~Callan, Jr., C.~Lovelace, C.~R.~Nappi and S.~A.~Yost,
  ``Adding Holes and Crosscaps to the Superstring,''
  Nucl.\ Phys.\ B {\bf 293} (1987) 83.

\bibitem{Cardy:1989ir}
  J.~L.~Cardy,
  ``Boundary Conditions, Fusion Rules and the Verlinde Formula,''
  Nucl.\ Phys.\ B {\bf 324} (1989) 581.

\bibitem{Ishibashi:1988kg}
  N.~Ishibashi,
  ``The Boundary and Crosscap States in Conformal Field Theories,''
  Mod.\ Phys.\ Lett.\ A {\bf 4} (1989) 251.

\bibitem{Cardy:1991tv}
  J.~L.~Cardy and D.~C.~Lewellen,
  ``Bulk and boundary operators in conformal field theory,''
  Phys.\ Lett.\ B {\bf 259} (1991) 274.

\bibitem{Pradisi:1996yd}
  G.~Pradisi, A.~Sagnotti and Y.~.S.~Stanev,
   ``Completeness conditions for boundary operators in 2-D conformal field theory,''
  Phys.\ Lett.\ B {\bf 381} (1996) 97
  [hep-th/9603097];
G.~Pradisi, A.~Sagnotti and Y.~.S.~Stanev,
  ``The Open descendants of nondiagonal SU(2) WZW models,''
  Phys.\ Lett.\ B {\bf 356} (1995) 230
  [hep-th/9506014];
G.~Pradisi, A.~Sagnotti and Y.~S.~Stanev,
  ``Planar duality in SU(2) WZW models,''
  Phys.\ Lett.\ B {\bf 354} (1995) 279
  [hep-th/9503207].

\bibitem{Fuchs:1997kt}
  J.~Fuchs and C.~Schweigert,
  ``A Classifying algebra for boundary conditions,''
  Phys.\ Lett.\ B {\bf 414} (1997) 251
  [hep-th/9708141].

\bibitem{Fuchs:2000cm}
  J.~Fuchs, L.~R.~Huiszoon, A.~N.~Schellekens, C.~Schweigert and J.~Walcher,
  ``Boundaries, crosscaps and simple currents,''
  Phys.\ Lett.\ B {\bf 495} (2000) 427
  [hep-th/0007174].

\bibitem{Walton:1987bu}
  M.~A.~Walton,
  ``The Heterotic String on the Simplest Calabi-yau Manifold and Its Orbifold Limits,''
  Phys.\ Rev.\ D {\bf 37} (1988) 377.

\bibitem{Wendland:2000ry}
  K.~Wendland,
  ``Consistency of orbifold conformal field theories on K3,''
  Adv.\ Theor.\ Math.\ Phys.\  {\bf 5} (2002) 429
  [hep-th/0010281].

\bibitem{Blumenhagen:2002wn}
  R.~Blumenhagen, V.~Braun, B.~Kors and D.~Lust,
  ``Orientifolds of K3 and Calabi-Yau manifolds with intersecting D-branes,''
  JHEP {\bf 0207} (2002) 026
  [hep-th/0206038].

\bibitem{Serre}
See, {\it e.g.}, J. P. Serre, ``A Course in Arithmetic'', Graduate Texts in Mathematics, Volume 7, Springer-Verlag, 1973.

\bibitem{Angelantonj:1999xf}
  C.~Angelantonj and R.~Blumenhagen,
  ``Discrete deformations in type I vacua,''
  Phys.\ Lett.\ B {\bf 473} (2000) 86
  [hep-th/9911190].

\bibitem{Pradisi:1999ii}
  G.~Pradisi,
  ``Type I vacua from diagonal Z(3) orbifolds,''
  Nucl.\ Phys.\ B {\bf 575} (2000) 134
  [hep-th/9912218].


\bibitem{Hull:2004in}
  C.~M.~Hull,
  ``A geometry for non-geometric string backgrounds,''
  JHEP {\bf 0510} (2005) 065
  [arXiv:hep-th/0406102];
  {\sl ibidem},
  ``Doubled geometry and T-folds,''
  JHEP {\bf 0707} (2007) 080
  [arXiv:hep-th/0605149];
  C.~M.~Hull and R.~A.~Reid-Edwards,
  ``Gauge Symmetry, T-Duality and Doubled Geometry,''
  JHEP {\bf 0808} (2008) 043
  [arXiv:0711.4818 [hep-th]].

\bibitem{deBoer:2012ma}
  J.~de Boer and M.~Shigemori,
  ``Exotic Branes in String Theory,''
  Phys.\ Rept.\  {\bf 532} (2013) 65
  [arXiv:1209.6056 [hep-th]].

\end{thebibliography}
\end{document}